\documentclass{article}
\usepackage{epsf}
\begin{document}

\title{ Large $Nc$ Behavior of Light Scalar Meson Nonet Revisited}
\author{Masayuki \sc Uehara\footnote{E-mail: ueharam@cc.saga-u.ac.jp}\\
Takagise-Nishi 2-10-17, Saga 840-0921, Japan}
\date{\today }
\maketitle
\begin{abstract}
We study whether light scalar meson nonet can survive in large 
values of $N_c$ through observing how scattering amplitudes behave as 
$N_c$ increases from 3 within a unitarized chiral approach. 
We obtain the result that vector mesons such as $\rho$ and $K^*$ survive as 
narrow width resonances, but all of the scalar meson nonet below 1 GeV fade 
out as $N_c$ exceeds a rather small number about 6.
 \end{abstract}
\def\beq{\begin{equation}}
\def\eeq{\end{equation}}
\def\beqa{\begin{eqnarray}}
\def\eeqa{\end{eqnarray}}
\def\fpi{f_\pi}
\def\mpi{m_\pi}
\def\mK{m_K}
\def\noeq{\nonumber} 
\def\mib#1{\mbox{\boldmath{$#1$}}}
\def\vt{\mib{t}}
\def\vJ{\mib{J}}
\def\re{{\rm Re}} \def\im{{\rm Im}}

\section{Introduction}
It has been shown by  Pel\'aez~\cite{Pelaez} that the 
complex poles corresponding to the  $\rho$ and $K^*$ mesons move toward the 
real axis  on the second Riemann sheet as a number of colors, $N_C$, becomes 
large, and that in contrast to the vector mesons those corresponding to the 
$\sigma$ and $\kappa$ states move away from the real axis, within  
the inverse amplitude method~(IAM) using full $O(p^4)$ amplitudes of the 
chiral perturbation theory~(ChPT)\cite{DP,Hanna,Guerr,GNP,PGN}. 

Stimulated by this work, we have calculated how physical quantities such as 
phase shifts and cross sections behave on the real axis of the physical sheet 
as $N_c$ increases from 3 to some finite values\cite{MU08}. In this 
calculation we adopt an approximate version of the two-channel IAM 
developed by Oller-Oset-Pel\'aez~(OOP)\cite{OOP,MU04}, 
which we call the OOP version. We have obtained the result that while the 
vector mesons become narrower resonances, the light scalar meson nonet 
including  the $f_0(980)$ and $a_0(980)$ states fade out as $N_c$ exceeds 
about 6.   
This result has been confirmed by Pel\'aez in his new paper, excluding an 
exceptional case of the $a_0(980)$ state\cite{Pelaez3}.

The same issue have been discussed by Oller and Oset\cite{Oller} using a 
different model, in which chiral $O(p^2)$ amplitudes and possible  preexisting 
tree resonance poles are introduced as ingredients of the model.  Their criterion 
whether a meson is dynamical or not is that the partial wave amplitude of 
the model can reproduce the experimental data 
without a nearby preexisting pole.  The preexisting 
poles are assumed to survive in the large $N_c$ limit. They have concluded that 
while the $\rho$ and $K^*$ mesons need each preexisting pole, light scalar 
mesons, possibly except for the $f_0(980)$ state, do not necessarily need such 
poles.  We note, however, that if energies to be fitted by the model are 
restricted to 1.2 GeV, the preexisting pole is not needed for the $f_0$ 
state\cite{Oller1}.

Thus, the above observations are consistent with the common understanding 
that the members of the vector meson nonet including  $\rho$ and $K^*$ 
are  typical of $q\bar q$ mesons in large $N_c$ QCD\cite{tHooft,Witten}. 
On the other hand the behavior of scalar mesons is at variance with the nature 
of the $q\bar q$ mesons. 

In this paper we study again the behavior of two-meson scattering 
amplitudes when $N_c$ increases from 3 under some different conditions from 
previous work,  within the two-channel OOP version with $O(p^4)$ 
amplitudes given in Ref.~\cite{GNP}. The $O(p^4)$ amplitudes depend 
on values of the low energy constants~(LEC) of ChPT, denoted as $L_n$, 
which are to be determined phenomenologically so as to reproduce experimental 
data. It would be difficult, therefore, to discriminate the nature of resonances 
by studying the $N_c=3$ world alone. 
In order to know the nature of the low mass mesons, it will be very useful to 
study how light vector and scalar meson states behave and how complex 
poles of $f_0(980)$ and $a_0(980)$ move when we increase $N_c$ from 3 to 
some values.
Although we do not go far away from the real $N_c=3$ world 
with $SU(3)\times SU(3)$ chiral symmetry, we observe that the 
$\rho$ and $K^*$ meson survive as narrow width resonances, 
but the light scalar nonet fade out as $N_c$ exceeds a rather small number 
about 6. This is the same result as in the previous work\cite{MU08}.

In the next section the explicit $N_c$ dependence of the OOP amplitudes is 
given, the vector channel and scalar channel are discussed in section 2 and 3, 
and the conclusions and discussion is given in the last section. 

\section{$N_c$ dependence of the OOP amplitudes}
In order to carry out the study we have to find the explicit $N_c$ dependence 
of the scattering amplitudes. The amplitudes in ChPT have an explicit 
$N_c$ dependence through the pion decay constant and the LECs. 
Since the pion decay constant $f_\pi$ is of 
$O(N_c^{1/2})$ and the LECs, $L_1$, $L_2$, $L_3$, $L_5$ and $L_8$ are to be 
of $O(N_c)$, but $2L_1-L_2$,  $L_4$, $L_6$ and $L_7$ are of 
$O(1)$,\cite{GL, PerisRaf, Siklody} we put
\beqa
L_n(N_c)&=& {\widehat{L_n}}\cdot\frac{N_c}{3}+\Delta L_n, \\
f_\pi^2(N_c)&=&\widehat{f_\pi^2}\cdot\frac{N_c}{3},
\eeqa
where $\widehat{L_n}$ satisfy the relations, $2\widehat{L_1}-
\widehat{L_2}=\widehat{L_4}=\widehat{L_6}=\widehat{L_7}=0$, and 
$\Delta L_n$ are of $O(1)$. Thus, we have 
\begin{equation}
\frac{L_n}{f_\pi^2}=\frac{\widehat{L_n}}{\widehat{f_\pi^2}}
+\frac{\Delta L_n}{\widehat{f_\pi^2}}\cdot\frac{3}{N_c} 
\label{Nc}
\end{equation}
with $\widehat{f_\pi}=93$ MeV. We also assume that the meson decay 
constants are the same and equal to the pion decay constant 
$\widehat{f_\pi}$ as in Ref.~\cite{GNP}.  

The ingredients of the IAM consist of amplitudes of chiral order $O(p^2)$ 
and $O(p^4)$ of ChPT. An $O(p^2)$ amplitude, denoted by  $T^{(2)}(s,t,u)$, has 
a form of a linear function of $s,~t,~u$ divided by $f_\pi^2$, and it is of 
$O(N_c^{-1})$. A polynomial term of the latter amplitudes, denoted by 
$T^{(4)}_{\rm poly}(s,t,u)$, is written as a sum of polynomial functions 
with the LECs as follows:
\begin{equation}
T^{(4)}_{\rm poly}(s,t,u)=\sum_{n=1,8}\frac{1}{f_\pi^2}
\left(\frac{L_n}{f_\pi^2}\right)P_n(s,t,u),
\end{equation}
where $P_n$ are quadratic functions of $s,~t,~u$ and meson mass 
squared. The polynomial term $T^{(4)}_{\rm poly}$ is 
of $O(N_c^{-1})$, because $L_n/f_\pi^2$ scales as $O(N_c^0)$ as seen in 
 Eq.(\ref{Nc}).  An $s$-channel loop term given by $t^{(2)}(s)J(s)t^{(2)}(s)$ is 
 of $O(N_c^{-2})$, where $J(s)$ is the one-loop function regularized as the 
$\overline{MS}-1$ scheme at the renormalization scale $\mu$\cite{GL}, 
where $t^{(2)}$ is a partial wave amplitude derived from $T^{(2)}(s,t,u)$. 
Similarly  $t$-  and  $u$-channel loop terms and tadpole terms are of 
$O(N_c^{-2})$, which are ignored in the OOP version. 
Thus, the OOP version is expected to be more valid as $N_c$ becomes larger. 
The $s$-channel loop terms are indispensable to realize unitarity, although 
they are of $O(N_c^{-2})$. This difference of the $N_c$ dependence produces the 
different behavior of the amplitudes when $N_c$ becomes large.  
 \begin{table}[h]
\label{tab:Ln}
\begin{center}
\begin{tabular}{|c|r|r|r|r|r|r|}\hline
&$ L_1$&$ L_2$&$ L_3$&$ L_5$&$ L_7$&$ L_8$ \\ \hline
Large $N_c$ &0.81&1.62&$-4.24$&1.21&$0$&0.60\\ \hline
Our $\widehat{L_n}$&0.70&1.40&$-3.20$&1.50&$0$&0.71\\ \hline
Our $\Delta L_n$&$0$& $-0.10$&$0$
&$0$&$-0.25$& $0$\\ \hline
\end{tabular}
\caption{$\ L_n\times 10^3$}
\end{center}
\end{table}
Our set of the LECs used in this work are determined at the renormalization 
scale $\mu=900$ MeV so as to reproduce experimental phase shifts qualitatively 
up to about 1.2 GeV. We note that the IAM and OOP amplitudes contain the LECs 
non-linearly and the fitting region is extended to higher energies.  
Our sets of $\widehat{L_n}$ and $\Delta L_n$ are tabulated  
in Table I with the set of the large $N_c$ model,\cite{Peris,Golterman} which 
correspond to $\widehat{L_n}$.  The change of the renormalization scale 
affects values of $\Delta L_n$, but we do not consider the scale change 
explicitly because the $\Delta L_n$ terms fade out as $N_c$ increases. 

We emphasize that since the Large $N_c$ set can reproduce the low energy 
scattering behavior rather well even at $N_c=3$, except for unessential points, 
 the results are almost the same if we take Large $N_c$ set instead of Our 
 $\widehat{L_n}$ and put $\Delta L_n$ to the difference between our $L_n$ 
 and the Large $N_c$ set.\\

\section{Vector channels}
At first, we discuss the behavior of vector mesons in the single channel 
calculation. The mass of a vector resonance is controlled by the combination 
of LEC, $2L_1-L_2+L_3$, \cite{DP} which is present in the term 
$\re[t^{(2)}-t^{(4)}_{\rm poly}]$  of $O(N_c^{-1})$,  and 
the loop contribution to the real part is of $O(N_c^{-2})$, so that the mass stays 
at an almost constant value. This is the IAM expression substituting for the 
preexisting pole. On the other hand  the imaginary part of the loop 
term contributes to the width. The octet component of the isoscalar vector 
meson 
\begin{center}
\begin{figure}[h]
\epsfxsize=10 cm
\centerline{\epsfbox{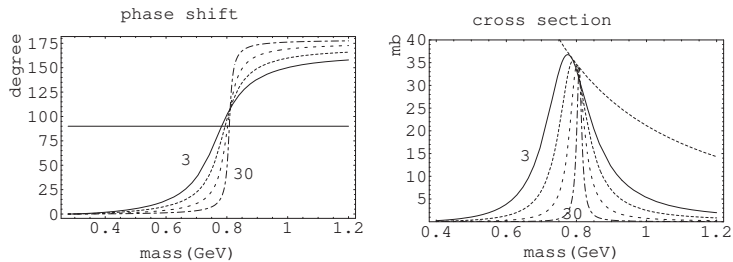}}
 \label{fig:rho}
\caption{$N_c$ dependence of phase shift (left) and 
 cross section (right) of the $\rho$ channel. Lines correspond to $N_c=3$, 5, 10 
  and 30 from the top to the bottom.}
\end{figure} 
\end{center}
\begin{center}
\begin{figure}[h]
\epsfxsize=10 cm
\centerline{\epsfbox{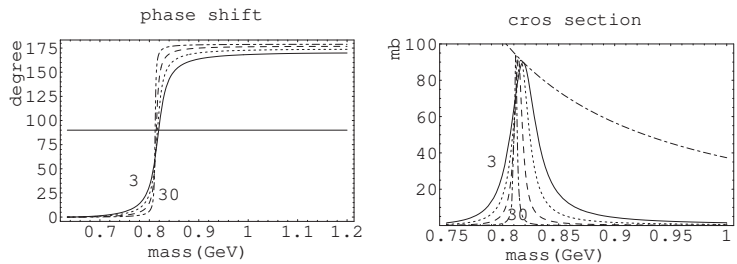}}
 \label{fig:Kstar}
\caption{$N_c$ dependence of phase shift (left) and 
 cross section (right) of the $K^*$ channel. Lines correspond to $N_c=3$, 5, 10 
 and 30 from the top to the bottom. }
\end{figure} 
\end{center}
also has  a constant mass and a residue of $O(N_c^{-1})$  below the 
$K\bar K$ threshold. Thus, the masses of the vector mesons  
are of $O(1)$, while the widths decrease as $O(N_c^{-1})$.

If we extend the calculation to the multi-channel IAM, 
the masses are almost unchanged, because the same combination of the LECs 
dominantly determine the masses. The $N_c$ dependence of the 
phase shifts and that of the cross sections of $\rho$ and $K^*$ are shown in 
Figs. 1 and 2.  The same behavior is observed in the case of the Large 
$N_c$ set, though the obtained  masses are too small because of too large
value of $|L_3|$.\footnote{The constraint $2L_1=L_2$ induces a large  
cancellation in $\det[t^{(2)}-t^{(4)}-{\rm loop}]$ even at $N_c=3$, so that the 
resultant amplitude gets a large uncertainty. But the narrowing width  
with increasing values of $N_c$ remains valid. }

Thus, we can conclude that the vector mesons 
described by the OOP version have the nature consistent with the $q\bar q$  
mesons. 

\section{Scalar channels}
\subsection{$(I,~J)=(0,~0)$}
This channel contains the controversial $\sigma(600)$ and the $f_0(980)$ 
states.  Using the two-channel IAM consisting of the $\pi\pi$ 
and $K\bar K$ channels,  we can reproduce experimental data fairly well 
below 1.2 GeV.
The  $N_c$ dependence of the phase shift and the cross section are shown in 
Fig. 3, where $N_c$ increases from 3 to 15. 
In contrast to the vector channel we observe that the 
phase shift becomes flat and the cross section fades out as $N_c$ becomes 
large; 
the sharp rise of the phase shift near the $K\bar K$ threshold and the large 
bump of the cross section near 
\begin{center}
\begin{figure}[h]
\epsfxsize=10 cm
\centerline{\epsfbox{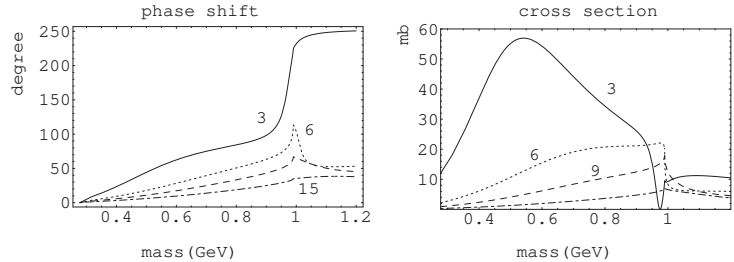}}
 \label{fig:fig00}
\caption{$N_c$ dependence of the phase shift~(left) and the 
 cross section~(right) of the (0,0) channel. Solid, dotted, dot-dot-dashed 
 and dashed lines are for $Nc=$3, 6, 9 and 15 respectively.}
\end{figure} 
\end{center}
\begin{center}
\begin{figure}[h]
\epsfxsize=10 cm
\centerline{\epsfbox{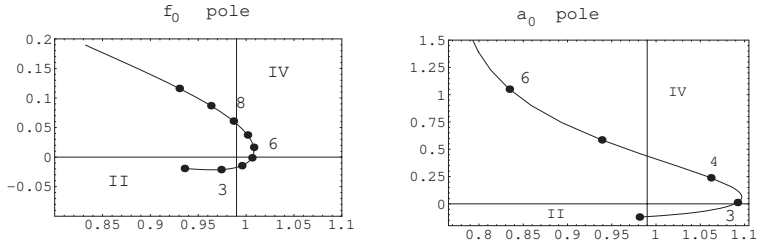}}
 \label{fig:pole}
\caption{$N_c$ dependence of the $f_0(980)$ pole~(left) and $a_0(980)$ 
pole~(right). Both of the poles wind around the branch point at $K\bar K$ 
threshold to go upward on the IV sheet. }
\end{figure} 
\end{center}
500 MeV seen at $N_c=3$  to 5  
disappears at $N_c=6$, and then the phase shift and the cross section 
become almost flat and fade out. Similar drastic change in the $N_c$ 
dependence has also been observed in Ref.~\cite{Harada}, though it 
is in different context.

The $f_0(980)$ pole exists at $(975-22i)$ MeV at $N_c=3$. Where does the
pole go as $N_c$ increases ? We approximately 
calculate the pole position by expanding the amplitude in powers of $k_2$
 up to the first order, where $k_2$ is the momentum of the $K\bar K$ channel.  
 We  observe that the pole moves into the upper half plane of the IV sheet 
 from the lower half plane of the II sheet, winding around the branch point at 
$K\bar K$ threshold, and goes away from the real axis as shown in the left 
side of Fig. 4.  Pole positions at larger $N_c$ cannot be 
reliable owing to the rough approximation, but this behavior would remain 
intact. 

We briefly comment on effects of adding the $\eta\eta$ channel to the OOP 
amplitude. If we include the $\eta\eta$ channel, we find that 
both real and imaginary part of $\det[t^{(2)}-t^{(4)}-{\rm loop terms}]$ 
develop zeros at almost the same point near 770 MeV even at $N_c=3$ for 
a wide range of the LEC sets. Such unreasonable behavior is also seen in the 
isospinor channel with the $\pi K$ and $\eta K$ channels as will be noted. 
These zeros do not violate unitarity, but give very unreasonable behavior to 
the amplitude, that is  too narrow resonant behavior with almost zero width.  
It should be noted, however, that the behavior of the amplitude, excluding a 
narrow strip including the zeros, is almost the same with the behavior of 
the two-channel model except for a shallow dip of the inelasticity at the 
$\eta\eta$ threshold\cite{Oller1}. 
The unpleasant behavior is not reported in the calculations with full 
$T^{(4)}$\cite{GNP,Pelaez3}. If we eliminate the unreasonable behavior, 
 the fading-out tendency as increasing $N_c$ remain valid, though the 
fading-out  occurs at larger $N_c$ owing to the additional $\eta$ loop 
contributions. 

\subsection{$(I,~J)=(1,~0)$}
This channel contains the $a_0(980)$ state and appears  as a cusp-like sharp 
peak as seen in Fig. 5. 
\begin{center}
\begin{figure}[h]
\epsfxsize=10 cm
\centerline{\epsfbox{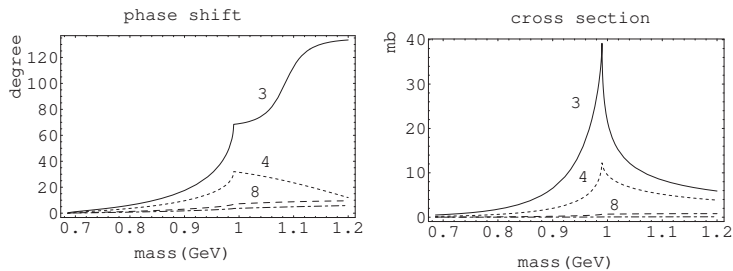}}
 \label{fig:fig(10)}
\caption{$N_c$ dependence of the phase shift~(left) and the
 cross section~(right) of the (1,0) channel. $N_c=3$ , 4, 8 and 12 from the top 
 to bottom.}
\end{figure} 
\end{center}
The rising phase shift after the cusp bends down and becomes to a flat curve, 
and the cross section having  a sharp peak fades out as $N_c$ increases.
The pole appears at $(1091-17i)$ MeV on the II sheet at $N_c=3$, and it
moves from the II sheet to the IV sheet, and leaves rapidly the real axis as 
shown in the left side of Fig.~4 as $N_c$ increases.  
We note that the real parts of the poles of the 
$f_0(980)$ and $a_0(980)$ states are not necessarily degenerate with each 
other, though both of the peaks of the mass distribution at $N_c=3$ appear near 
the $K \bar K$ threshold owing to the cusp behavior of the $a_0(980)$ state.

\subsection{$(I,~J)=(1/2,~0)$}
The fading-out behavior of the phase shift and cross section with increasing 
$N_c$ is the same as that in the channels discussed above.  As stated before  
 there appears an artifact zero near 750 MeV originated from the 
 $\eta K\to \eta K$ component in the OOP version used in this work.  
 So if we eliminate the zero by an interpolation method, we observe that 
 the results by  the two channel model are  almost the same as those by the 
 single channel calculation by virtue of the weak coupling between the $\pi K$ 
 and  $\eta K$ channels. 
 The calculations making use of the full $T^{(4)}$ do not give such 
 an unwanted zero~\cite{GNP}. 
\begin{center}
\begin{figure}[h]
\epsfxsize=10 cm
\centerline{\epsfbox{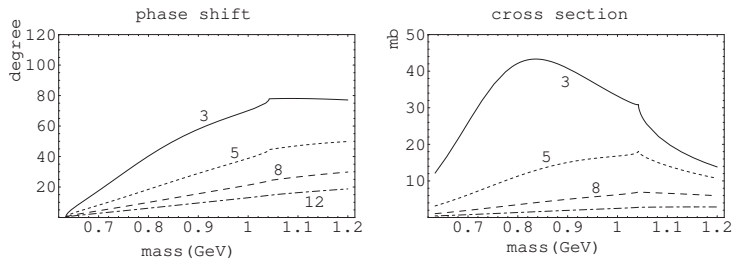}}
 \label{fig:fig(1/20)}
\caption{$N_c$ dependence of phase shift~(left) and 
 cross section~(right) of the (1/2,0) channel. $N_c=3,$ 5, 8 and 12 from the 
 above  to bottom.}
\end{figure} 
\end{center}

\section{Concluding remarks}
We have calculated the $N_c$ dependence of the vector and scalar channels 
stating from $N_c=3$ to finite values, 30 for the vector channel and 12 or 
15 for the scalar channel within the approximate IAM under the 
$N_c$ dependence of $L_n/f_\pi^2$ given by Eq.(\ref{Nc}) and 
$f_\pi(N_c)=\sqrt{N_c/3}\times f_\pi(3)$. And we have observed that 
the vector mesons survive as sharper resonances at almost the same position, 
the resonant structures of the scalar channels fade out at rather low values 
of $N_c$ near 5 or 6. By extending the observation we are led to conclude  
that the vector meson nonet has the nature 
consistent with the $q\bar q$ mesons in large $N_c$ QCD, but  
the light scalar meson nonet cannot survive in large $N_c$ and 
then cannot have the nature of the $q\bar q$ mesons. This conclusion   
is the same as obtained by Pel\'aez, excluding an exceptional case of the 
$a_0(980)$ state,\cite{Pelaez3} and it is also consistent with the results 
obtained in Ref.~\cite{Oller,Oller1}.
Our conclusion supports the arguments that the light scalar 
mesons are of the $K\bar K$ molecule\cite{Wein,Lohse,Janssen}, and  
of $q^2{\bar q}^2$ states\cite{Jaffe, Achasov,Kumano}. 

If the $f_0$ and $a_0$ states are composed of $(qs)(\bar q\bar s)$ state, 
where $q$ denotes $u$ and/or $d$ quark, the similarity 
between the $f_0$ and $a_0$ both in mass and generating mechanism is  
expected. However, the pole positions of the both states can be different 
from each other by  about 100 MeV or more, and the  generating mechanism 
of the $a_0(980)$ state would be different from that of $f_0$ state: 
The $a_0$ state is generated by the strong channel coupling between 
$\pi \eta$ and $K\bar K$ channels, but not as a bound state resonance like 
as the $f_0$ state, as seen in the exchange dynamics\cite{Wein,Janssen} and 
in the chiral loop dynamics\cite{OOBS,MU04}. There is also the argument that 
the $a_0(980)$ and $f_0(980)$ are not elementary particles within the 
hadronic dynamics\cite{Baru}. They have observed that the field 
 renormalization constants $Z$ of both states are close to 0 using the 
 propagators of existing models, and concluded that 
a simple $q\bar q$ or four quark assignment for the $a_0$ should be 
considered with caution and it is certainly questionable for the $f_0$. 
If the light scalar nonet are not $q\bar q$ mesons, we cannot include them 
into mass spectra in the low meson dominance hypothesis, 
because the low mesons are supposed to participate in narrow resonance 
towers in large $N_c$ limit of 
QCD\cite{Peris,Golterman,Ecker,Bijnens,Polyakov}. 

 Our conclusion strongly indicates that all of the light scalar 
nonet are  dynamical effects originating from unitarity, chiral symmetry 
and strong channel couplings. If the mesons in the scalar nonet are dominantly 
composed of hadronic or four quark component, but include $|q\bar q>_P$ 
with a small fraction as in Ref.~\cite{Close}, we could find out the small 
$|q\bar q>_P$ component by increasing $N_c$ in theoretical models, because 
the large hadronic or four quark component fades out and the $q\bar q$ 
component remains. At least,  our calculation within the two-channel OOP 
approximation does not indicates that such an intriguing change will occur 
in larger $N_c$ region.


\end{document}